# Students' Acceptance of Arduino Technology Integration in Student-Led Science Inquiry: Insights from the Technology Acceptance Model

Seok-Hyun Ga, Chun-Yen Chang, Sonya Martin

*Abstract*—This study examines high school students' acceptance of Arduino technology in a student-led, inquiry-based science class, using the extended Technology Acceptance Model (TAM2) as a guiding framework. Through qualitative analysis of interviews and classroom observations, we explored how students perceived Arduino's usefulness and ease of use. Going beyond traditional quantitative TAM studies, this qualitative TAM research provides a nuanced, in-depth understanding of the contextual factors shaping technology acceptance. Key findings reveal that acceptance was driven not only by instrumental factors like job relevance and output quality but also by the unique sociocultural context of the Korean education system, where technology use was perceived as valuable for university admissions (subjective norm and image). Critically, unlike earlier research that emphasized programming challenges, participants in this study found Arduino accessible and intuitive, thanks to integrated visual block-coding tools. These findings highlight the importance of both technological design and pedagogical support in shaping students' experiences. Implications for science curriculum design, teacher preparation, and equitable technology integration in secondary education are discussed.

This work was supported by the National Research Foundation of Korea (Grant No. 2023S1A5B5A16080581), the National Science and Technology Council, Taiwan (Grant No. NSTC-114-2410-H-003-027-MY3), and the Institute for Research Excellence in Learning Sciences of National Taiwan Normal University (NTNU) from the Featured Areas Research Center Program within the framework of the Higher Education Sprout Project by the Ministry of Education (MOE) in Taiwan.

This work involved human subjects or animals in its research. Approval of all ethical and experimental procedures and protocols was granted by ethics committee of the Seoul National University under Application No. IRB No. 2309/003-027.

Seok-Hyun Ga is with the Center for Educational Research, Seoul National University, Seoul 08826, Republic of Korea, and the Graduate Institute of Science Education, National Taiwan Normal University, Taipei 10610, Taiwan (e-mail: shga@snu.ac.kr).

Chun-Yen Chang is with the Institute for Research Excellence in Learning Sciences, Graduate Institute of Science Education, and Department of Earth Sciences, National Taiwan Normal University, Taipei 11677, Taiwan; the Department of Biology, Universitas Negeri Malang, Malang 65145, Indonesia; and the Graduate School of Education, Chung Yuan Christian University, Taoyuan 32023, Taiwan (email: changcy@ntnu.edu.tw).

Sonya N. Martin is with the Department of Biology Education, Seoul National University, Seoul 08826, Republic of Korea (email: sonya_martin@fastmail.com).

## I. INTRODUCTION

AS technology becomes increasingly embedded in everyday life, educational systems strive to prepare students with the skills needed for future success, particularly in science, technology, engineering, and mathematics (STEM) disciplines. Among emerging educational technologies, physical computing platforms, especially Arduino, have gained prominence in science education for their ability to make scientific inquiry more accessible, personalized, and engaging. Arduino is an open-source electronics platform that integrates affordable hardware with intuitive software, enabling students to build custom scientific instruments by connecting sensors and actuators [1], [2], [3]. This approach empowers learners to design and carry out experiments aligned with their own research interests and investigative goals. Given these advantages, it is important to understand how students perceive and adopt Arduino technology within authentic educational settings, particularly through the lens of established models such as the Technology Acceptance Model (TAM).

Scientific inquiry is a foundational element of contemporary science education, valued for its role in fostering students' critical thinking and problem-solving abilities [4]. Schools increasingly encourage student participation in hands-on inquiry through both observational and experimental activities. Chin *et al.* [5] emphasize that authentic inquiry, in which students generate questions and design procedures, supports a deeper understanding of scientific concepts. However, despite its recognized benefits, student-led investigations often face practical challenges. A key barrier is the lack of diverse, adaptable, and affordable measurement tools in school settings [3], which limits students' ability to engage fully in data-driven inquiry. To address this, science educators must provide both the tools and pedagogical supports necessary to facilitate meaningful scientific investigations.

Physical computing platforms such as Arduino help to overcome these limitations by offering low-cost, flexible solutions for building personalized data-collection systems. These tools support authentic inquiry by enabling customized experimental design and real-time data acquisition [6], [7]. Additionally, platforms like Arduino encourage collaborative engagement by enabling students to share and analyze data with peers, fostering inclusive participation in scientific practices [8], [9].

*A. Addressing Challenges Through Technological Innovations*

In the investigation of Arduino use in classroom science, text-based coding was identified as a significant barrier to adoption [3]. To address this issue, user-friendly visual programming tools such as mBlock, along with specialized extensions like the Gaduino package, have been developed to lower the entry barrier for students and teachers [2].

mBlock simplifies Arduino programming by using drag-and-drop visual blocks, which reduces the complexity of syntax and minimizes errors typically associated with traditional coding. However, mBlock alone may not meet the specific sensor and data collection needs of scientific experiments. The Gaduino package extends mBlock's capabilities with additional blocks tailored for scientific applications, including controls for displays, communication modules, and sensors that measure environmental variables such as gas concentration and fine particulate matter. Together, these innovations reduce the cognitive load of programming, enhance ease of use, and allow students to focus more deeply on scientific investigation.

These technological supports align with key constructs in the Technology Acceptance Model (TAM), a widely used framework for understanding how users' perceptions of a technology's usefulness and ease of use influence their willingness to adopt and apply it. In the context of inquiry-based learning, where students take on active roles in designing investigations, their acceptance of classroom technologies becomes particularly important. By enabling student-centered experimentation without the burden of complex programming, mBlock and Gaduino have the potential to increase students' confidence and motivation to use Arduino.

This study draws on TAM to qualitatively explore how high school students' experience and evaluate Arduino within a student-led inquiry project. While numerous studies have applied TAM/TAM2 to educational technologies, they have often relied on quantitative survey methods, which reveal correlations but fail to capture the rich, contextualized 'why' behind students' perceptions. Furthermore, much of the existing literature on Arduino in education points to significant challenges related to text-based programming, often framing it as a barrier for novice learners [10].

This leaves a gap in our understanding of two key areas. First, it is unclear how newer pedagogical tools, like visual coding environments, might change this challenging dynamic for students. Second, few studies have qualitatively explored what 'usefulness' truly means to high school students in competitive academic environments where factors like social pressures and career goals are highly influential. Therefore, this study adopts a qualitative approach to address these gaps. Specifically, we examine how students' perceptions of usefulness and ease of use shape their acceptance and identify instructional strategies that may enhance the integration of Arduino in secondary science education.

*B. Research Questions*

Recognizing the importance of student perceptions in technology adoption, this study investigates the following questions:
1) What aspects of Arduino's perceived usefulness influenced students' acceptance of the technology for conducting scientific inquiry?
2) How did students' perceptions of Arduino's ease of use, particularly through visual block-coding interfaces, impact their acceptance of the technology in a student-led inquiry context?
3) What pedagogical strategies or supports could educators employ to enhance students' acceptance and effective use of Arduino in inquiry-based science education settings?

By addressing these questions, the study aims to inform educators, curriculum developers, and policymakers about effective strategies for integrating Arduino in science classrooms. The findings will support efforts to deepen students' scientific understanding, improve problem-solving competencies, and increase engagement through meaningful use of educational technologies.

2. THEORETICAL BACKGROUND

The Technology Acceptance Model (TAM), developed by Davis [11], offers a foundational framework for examining how users accept and adopt new technologies. As shown in Figure 1 below, TAM posits that users' behavioral intentions and eventual use of a technology are shaped by two core constructs: perceived usefulness and perceived ease of use.

Perceived usefulness refers to "the degree to which a person believes that using a particular system would enhance his or her job performance" [11, p. 320]. In educational settings, this concept encompasses students' expectations that a tool will enhance their learning effectiveness or efficiency. Perceived ease of use denotes "the degree to which a person believes that using a particular system would be free of effort" [11, p. 320]. Technologies perceived as simple and intuitive are generally more likely to be adopted, as they reduce resistance associated with technical complexity.

Although TAM was initially developed within organizational and business settings, its application has broadened to include diverse fields such as education, healthcare, and consumer technologies. Educational researchers have adopted TAM to study various technologies and user populations, often confirming its core constructs as predictors of technology acceptance. Numerous empirical studies have confirmed the utility of its core constructs in predicting technology acceptance and guiding implementation strategies (Açıkgül & Şad, 2021; Alasmari & Zhang, 2019; Chahal & Rani, 2022; Ha & Stoel, 2009; Hsu & Lin, 2022; Lin & Kim, 2016; Lin & Yeh, 2019; Rauniar et al., 2014; Song & Kong, 2017).

[**Insert Figure 1 here: The original Technology

**Acceptance Model (adapted from [21, p. 984])]**

While TAM provides a useful foundation for understanding core beliefs about technology, it does not fully account for the social and contextual variables that often influence students in classroom settings. To address this limitation, Venkatesh et al. [22] introduced an expanded version, TAM2, which incorporates additional constructs that reflect both social influences and cognitive instrumental processes (see Figure 2). These constructs include subjective norm, voluntariness, image, job relevance, output quality, and result demonstrability. Each of these helps to explain how external factors shape perceptions of usefulness and ease of use in more nuanced ways.

In TAM2, subjective norm captures the influence of perceived social pressure to adopt a technology, while voluntariness refers to the extent to which individuals view technology use as a choice. Image relates to the perceived enhancement of social standing through technology use. The instrumental components, including job relevance, output quality, and result demonstrability, reflect how well the technology aligns with users' goals, the quality of its performance, and the visibility of its benefits.

[**Insert Figure 2 here: Extended Technology Acceptance Model 2** (adapted from [22])]

In the present study, both TAM and TAM2 provide a conceptual lens for examining high school students' engagement with Arduino in an inquiry-based science class. TAM offers a foundation for understanding how perceived usefulness and ease of use shape students' attitudes toward using Arduino. TAM2 extends this foundation by accounting for factors such as peer influence, relevance to academic goals, and the clarity of technological benefits. These constructs are particularly important in educational contexts where student motivation and classroom dynamics influence technology adoption.

The constructs emphasized in TAM2 also align closely with the study's research questions. Perceived usefulness corresponds with the first research question, which explores students' evaluations of Arduino's value for scientific inquiry. Perceived ease of use informs the second question, which focuses on students' experiences with visual programming interfaces. The broader TAM2 framework, including instructional support and social context, contributes to the third question concerning effective strategies for promoting student acceptance of classroom technologies.

Research using TAM2 in educational contexts has provided valuable insights into how students and teachers adopt new tools. For example, Virdi and Mer [23] found that instructional quality played a significant role in shaping students' intentions to participate in online learning. Açıkgül and Şad [17] examined mobile learning acceptance among high school mathematics students using a related model, and Hsu and Lin [15] extended TAM to include motivational factors in mobile-assisted language learning. Additional studies by Chahal and Rani [16] and Mailizar, Almanthari and Maulina [24] have explored similar questions in higher education and among experienced teachers. While these studies affirm the flexibility of TAM2 across settings, they rely primarily on quantitative designs such as Likert-scale surveys, which may not capture the full range of student experiences.

Given the limitations of survey-based approaches, this study adopts a qualitative lens to explore students' acceptance of Arduino in greater depth. By collecting semi-structured interviews and observational field notes, we aim to identify the personal, social, and contextual factors that influence students' use of Arduino in student-led inquiry. Interviews allow students to reflect on their experiences in their own words, while classroom observations provide insight into peer interactions and instructional supports that shape engagement.

Through qualitative analysis informed by TAM2, this study seeks to uncover the complexities of how students interpret and adopt Arduino in science learning. In doing so, it contributes to a more nuanced understanding of technology acceptance in education and offers practical guidance for educators and curriculum designers working to integrate physical computing into inquiry-based instruction.

3. METHOD

*A. Research Context*

This study was conducted at a high school in Seoul, South Korea. In South Korea, secondary education comprises middle school (grades 7–9) and high school (grades 10–12). Students at the site could enroll in an after-school research course, Research and Education (R&E). Although scheduled by the school, the course was not part of the formal academic curriculum. The R&E class promoted student-led inquiry by encouraging participants to identify research questions based on their interests and to design and implement projects that developed scientific investigation skills. The classroom teacher noted that limited access to measurement equipment had previously hindered students' ability to pursue their research goals effectively.

To address this limitation, the researcher proposed integrating Arduino, a microcontroller platform, into the course. Arduino was selected because of its affordability, accessibility, and compatibility with various sensors, making it a practical solution for creating customized tools. Compared to other physical computing options, Arduino offered flexibility and broad availability, allowing students to develop more diverse and complex experimental setups tailored to their research questions.

Student recruitment for the R&E class was supported by the teacher, who introduced the opportunity during regular class time and via a school bulletin board. Six students voluntarily enrolled, and all were accepted. The participants were second-year high school students, equivalent to Grade 11 in the K to 12 system. Five students were enrolled in the natural sciences track, known as *i-gwa* (이과), which emphasizes science and mathematics. One student came from the humanities and social sciences track, or *mun-gwa* (문과), which focuses on literature, history, and related subjects. In South Korea, high schools typically allow students to choose from academic tracks aligned with their intended university major or career goals, including science, humanities, and occasionally the arts or physical education.

The participating teacher held a degree in Earth Science Education and had seven years of teaching experience. At the time of the study, the teacher taught Earth Science I and II, in accordance with the Korean 2015 National Curriculum for Primary and Secondary Schools. However, the teacher had no prior experience with Arduino or physical computing platforms.

*B. Session Overview*

The R&E class took place over nine sessions (see TABLE 1), each lasting 90 minutes. The first session introduced students to the course structure and purpose. Past student-led projects were shared to help participants generate ideas and begin formulating research topics. Since none of the students had previously conducted independent research, these examples served as valuable models.

In the second session, students were introduced to Arduino through a block-based coding environment, using the method developed by Authors 1 and 2. This approach used mBlock 5, a visual programming tool that simplifies coding and reduces the cognitive load associated with text-based programming.

The third session focused on concepts related to the Internet of Things. Students learned how to use an IoT platform to transfer data from Arduino devices to cloud-based systems for storage and visualization. This session also emphasized the relevance of data quality and demonstrability in real-world contexts. During this time, students refined their research ideas with guidance from the teacher and selected a final project topic.

In the fourth session, students constructed measurement devices aligned with their research questions. Those using Arduino developed and tested sensor-based tools, while others used equipment already available at school. This hands-on activity gave students the opportunity to evaluate Arduino's usefulness and ease of use in practical terms.

The fifth session was used to finalize project plans and outline implementation steps. Sessions six and seven were dedicated to project progress updates and iterative feedback from the teacher. In session eight, students formally presented their research outcomes. The ninth and final session was a structured reflection in which students evaluated their learning experiences and discussed the advantages and limitations of using Arduino.

[Insert TABLE 1 here: After-School Research Class Schedule]

*C. Data Collection*

The primary data sources for this study included observational field notes and semi-structured interviews. The researcher attended and observed all class sessions, recording information on lesson content, student challenges and successes, and interactions between students and the teacher. Field notes also captured contextual details, such as how students selected their research topics and how teacher support influenced their decisions.

Using insights from these observations, the researcher created a semi-structured interview protocol focused on students' perceptions of Arduino. Each interview lasted approximately 30 minutes and was conducted with five of the six participating students. One student opted out of the interview portion for personal reasons. All interviewed students used Arduino in their projects. Sample interview questions included:
1) What is the topic of your project?
2) Why did you select this particular topic?
3) What device or tools did you use to carry out your project?
4) Why did you choose to use Arduino instead of school-provided equipment?
5) What advantages did you perceive in using Arduino?
6) Was Arduino appropriate for achieving your research goals?
7) How did the data collected with Arduino help you address your research question?
8) Do you believe Arduino enabled you to collect high-quality data?
9) Would you use Arduino again in future projects?

All students and their legal guardians received full information about the research purpose, procedures, voluntary nature of participation, and their rights to withdraw at any time. Consent was obtained from both students and guardians. Pseudonyms were used to protect students' identities, and interview recordings were deleted after transcription to ensure confidentiality.

*D. Data Analysis*

This study adopted an interpretive qualitative approach to explore students' experiences with Arduino in a student-led science context. The process of data analysis is summarized in Figure 3.

[Insert Figure 3 here: Procedures of Data Analysis]

Interview transcripts were coded using the components of the Technology Acceptance Model 2 (TAM2; see TABLE 2): Perceived Ease of Use (PEU), Subjective Norm (SN), Image (IMG), Job Relevance (JOB), Output Quality (OQ), and Result Demonstrability (RD). The constructs of Experience and Voluntariness, while part of TAM2, were not coded independently, as they function as mediators influencing the impact of subjective norms on perceived usefulness and intention to use.

Two researchers independently coded the transcripts and compared their findings. Any discrepancies were resolved through discussion and consensus. Observational field notes were used to triangulate the data and enhance reliability. Following initial coding and theme identification, the researchers collaborated to synthesize the results and explore how each TAM2 component was reflected in the student experiences. A preliminary summary of findings was shared with the participating teacher for member-checking to confirm the accuracy and contextual relevance of the interpretations.

**[Insert TABLE 2 here: Components of the TAM2 Model Used in Coding]**

## 4. RESULTS AND DISCUSSION

*A. Students' Perceived Usefulness of Arduino*

This section addresses Research Question 1 by exploring how students perceived the usefulness of Arduino within their inquiry-based projects. Perceived usefulness, as defined in TAM2, refers to the belief that using a specific technology will enhance task performance. Students' responses reflected this belief across several interrelated constructs.

*1) Subjective Norm*: Students' decisions to use Arduino were influenced by perceived social expectations, particularly those related to university admissions and academic documentation. In the South Korean education system, the school record or *hakgyosaenghwal girokbu* (학교생활기록부), plays a significant role in university applications. Several students noted that they expected the inclusion of Arduino-based research in their records would make a favorable impression.

> *This activity will be written in my school record. I searched related science learning activities on the Internet, and there were quite a few Arduino activities, so I thought it would be fancy to use it for my research and appropriate to be included in the school life report.* (Student 1)

> *Honestly, I wanted to include content from other ICT disciplines in my school life report because the content of my report is so liberal arts oriented.* (Student 5)

Even though teachers emphasized that students were not required to use Arduino, its prominent introduction early in the course may have implicitly influenced students' decisions.

> *I knew I had to use Arduino for my project activities.* (Student 3)

These findings suggest that subjective norm, particularly the pressure to present advanced and relevant experiences in academic records, influenced students' willingness to adopt Arduino.

*2) Image*: Students associated Arduino with cutting-edge innovation, even when their understanding of its specific functions was limited. The association with high-tech fields contributed to a positive image of Arduino as a valuable and modern learning tool.

> *Arduino... Internet of Things... it's a bit of an image... a little bit of an image of a new technology... I don't know enough about it.* (Student 1)

> *I think Arduino could be useful if used well, so I'd like to learn about the new technology.* (Student 2)

> *I had heard a lot about Arduino before, but I was intrigued and curious. I wanted to learn about it someday.* (Student 4)

The perception of Arduino as novel and sophisticated appeared to increase students' confidence and interest. However, the findings also underscore the need for instructional clarity to ensure that decisions are based on informed understanding rather than reputation alone.

*3) Job Relevance*: Students evaluated Arduino's relevance in relation to their future career interests rather than to current coursework. While no participants reported direct curricular links, some saw the platform as aligned with their anticipated university studies.

> *I'm planning to study architecture in university, and IoT is really popular in that field.* (Student 1)

> *I'm kind of interested in coding and stuff like that... I'm kind of hoping for computer science as a future career.* (Student 2)

One student noted limited relevance but used Arduino to gain new experience.

> *My career aspirations are in life sciences... I just wanted to do something with an Arduino because I hadn't done much coding, and I wanted to see what it was like.* (Student 4)

These responses suggest that perceived job relevance may extend beyond immediate academic needs to broader career aspirations. Educators can enhance this relevance by explicitly linking Arduino applications to a wider array of fields.

*4) Output Quality*: Students generally expressed confidence in the reliability of the data collected with Arduino sensors.

> *I thought that measuring with an Arduino would be more accurate than other devices for carbon dioxide and particulate matter.* (Student 2)

> *I used Arduino because it gave me more accurate results than other measurement devices.* (Student 3)

However, some students noted limitations or anomalies in the data, pointing to a need for better understanding of sensor calibration and error margins.

> *I don't think the measurements are inaccurate... but there are some oddities in the graph, so I can't say it's completely accurate. However, I think it's good enough for my project.* (Student 4)

Although students viewed the output quality favorably, their judgments were largely intuitive. Introducing sensor datasheets and guiding students in evaluating technical specifications could improve their ability to assess measurement quality.

**5) Result Demonstrability**: Students valued Arduino's ability to collect, store, and visualize data over time, particularly through IoT platforms. This capability enhanced the perceived visibility and credibility of their findings.

> *With the Galaxy Watch or Apple Watch, you can only see the current heart rate, but with the Arduino, you can see the process of change.* (Student 3)

> *When you go into the IoT platform, you can see the measurements in real time, and you can also see historical data... I thought I could get a full day's worth of data with the Arduino.* (Student 4)

> *Illuminance is affected by sunlight... I chose an Arduino to better capture the changes over a few hours.* (Student 5)

However, some students questioned the reliability of indirect data visualization via the IoT platform.

> *You don't see the measurements right away on the device's display, but you can go into the computer (IoT platform) and see the numbers indirectly, so I was a little bit like, is this correct?* (Student 5)

Teachers could address such concerns by clarifying how data transmission and visualization work or connecting IoT platforms with direct LCD displays to provide clearer explanations about how sensor data is transmitted and visualized through the IoT platform. This may enhance students' trust in the data and improve perceived usefulness.

*4.2 Students' Perceived Ease of Use of Arduino*

Research Question 2 (RQ2 examined students' perceptions of how easy it was to use Arduino. In contrast to prior research that identified Arduino as challenging due to its reliance on text-based coding, all students in this study reported that they found the platform easy to use. '

> *It was not difficult for me to use it on my project.* (Student 1)

> *I had never done it before and I wasn't sure if I would be good at it, but I think I did okay.* (Student 3)

Two key factors contributed to students' perceptions that Arduino was easy to use. First, the implementation of block-coding interface using mBlock significantly lowered the entry barrier by removing the complexity associated with syntax-based coding. Students were able to program Arduino devices through visual drag-and-drop blocks, which reduced cognitive load and made programming more accessible to those with no prior experience. Second, the Gaduino extension, developed specifically for scientific applications, allowed students to interact with sensors, displays, and data transmission functions using single, clearly labeled blocks. Tasks that would normally require writing multiple lines of code, such as initializing a gas sensor or uploading data to an IoT platform, could be completed with a single visual block. This streamlined design enabled students to focus more fully on scientific reasoning and experimental planning, rather than on debugging or mastering complex coding syntax.

Students' comments reflected a general sense of confidence and satisfaction with the technical aspects of their projects. This finding contrasts with earlier studies [2], [25], [26], [27], which described Arduino as difficult for students to use, often due to its reliance on traditional text-based programming environments. In the present study, however, the simplified coding tools appeared to shift students' attention away from programming challenges and toward meaningful engagement with inquiry tasks.

By minimizing technical barriers, mBlock and Gaduino together supported a learning environment in which students could direct their cognitive resources toward scientific exploration. These tools not only improved ease of use but also played a central role in shaping students' overall acceptance of Arduino technology.

> *The teacher actually taught me how to do everything, so I don't think I had any difficulties with it.* (Student 2)

> *I just follow what I've learned, what's so hard about it?* (Student 4)

> *I was going to use an Arduino to build a light meter, but there's enough variety in the course material to make it useful even for those who aren't interested.* (Student 5)

In addition to the design of the tools themselves, pedagogical support played a central role in shaping students' perceptions of Arduino's ease of use. Students frequently cited early instructional sessions and classroom materials as essential to their ability to use the platform effectively. One student commented that the teacher's explanations made it easy to follow along, while another explained that simply applying what was taught in class was sufficient to complete their project. Even when students worked with sensors not directly covered in the lessons, such as carbon dioxide or light sensors, they felt the foundational materials provided enough guidance to proceed with confidence.

Although the instruction centered on a pre-structured activity using particulate matter sensors, students applied the same principles to build devices with different sensors, including those for heart rate and gas detection. This flexibility suggests that initial instruction was well-scaffolded. However, it also highlights the need for expanded materials that explain sensor characteristics and applications in greater depth. A deeper understanding of how sensors function, including how to interpret specifications like range and resolution, would further support students in adapting the technology to new contexts and research questions.

When asked whether they would use Arduino again in the future, all five interviewed students responded affirmatively. Their reasons reflected multiple TAM2 constructs. For some, subjective norm and image played a role, as shown in the comment, "*To enrich my school life record*" (Student 1). Others pointed to job relevance, with one student stating, "*Because I am very interested in the computer engineering field*" (Student 2). Students also cited perceived usefulness, both in terms of curiosity and functionality. One noted, "*I felt curious seeing another student use the heart rate sensor. I want to try using various other sensors too*" (Student 3), while another observed, "*Because a lot of data can accumulate even if left alone*" (Student 5). Ease of use was also mentioned, with a student saying, "*Because it's convenient since it measures well just by plugging it in*" (Student 4).

These varied responses demonstrate the multidimensional nature of technology acceptance. While some students were motivated by social recognition or career alignment, others valued the platform's convenience, data capabilities, or their own exploratory interests. This convergence of factors suggests that students' positive experiences with Arduino, supported by accessible tools and targeted instruction, may contribute to their sustained interest in using educational technologies during future inquiry activities.

*4.3 Pedagogical Strategies and Supports*

To address Research Question 3 (RQ3), this section synthesizes findings related to students' perceptions of Arduino's usefulness and ease of use to identify instructional strategies that may enhance its integration in science classrooms. Students' adoption of Arduino was shaped by both the design of the technology and the social and instructional context in which it was introduced. Their decisions reflected multiple TAM2 constructs, including perceived usefulness, ease of use, subjective norm, image, job relevance, and result demonstrability.

The data suggest that educators play a pivotal role in facilitating students' acceptance of classroom technologies. One key area of focus is clarifying how Arduino-based systems collect and transmit data, especially when using cloud-based Internet of Things platforms. Several students expressed uncertainty about the accuracy of data they could not observe directly on a local display. To address this concern, teachers can provide explicit explanations of how sensor data is processed and visualized and, where possible, integrate display components such as LCD screens into student devices. Making data outputs more immediate and transparent may improve students' trust in the results and increase their perceived confidence in using the system.

Another instructional consideration involves helping students evaluate sensor quality. While students appreciated the ability to collect accurate data, their judgments were often based on intuitive assessments or previous exposure rather than on objective criteria. Many selected sensors without fully understanding their specifications or capabilities. Introducing sensor datasheets and guiding students through concepts such as sensitivity, range, and resolution would strengthen their ability to select and apply components that align with their scientific questions. This form of guidance not only supports inquiry design but also fosters critical thinking about instrumentation and measurement.

Finally, the scope of instructional materials should be expanded to include a broader range of sensor types and scientific applications. While the pre-designed activities served as a strong foundation, they focused primarily on a single sensor type. Students were able to extend their learning to new components, but this process required them to infer connections not explicitly addressed in the lessons. Supplementing hands-on activities with conceptual explanations of how different sensors work and when to use them would help students generalize their knowledge and develop greater autonomy in project design.

Together, these strategies can enhance students' confidence and ability to engage with Arduino technology in meaningful ways. By supporting both technical competence and scientific reasoning, educators can help ensure that students' initial positive experiences with Arduino lead to continued engagement and deeper learning in inquiry-based science education.

5. CONCLUSIONS

This study examined the factors influencing high school students' acceptance of Arduino technology within the context of inquiry-based science education, using the Technology Acceptance Model 2 (TAM2) as a guiding framework. The findings demonstrated overall positive acceptance among participants, with all students selecting Arduino as the primary measurement device for their projects and reporting it as both useful and easy to use.

In response to the first research question concerning perceived usefulness, several influencing factors emerged. Students believed that working with Arduino would strengthen their school life records, particularly in ways advantageous for university admissions. Arduino's image as a modern and sophisticated technology further supported its appeal. Some students also viewed it as aligned with their future academic or career interests. Additional contributors to perceived usefulness included students' confidence in the quality of the data collected and their ability to track data changes over time through visualization on Internet of Things (IoT) platform.

With regard to the second research question on ease of use, students consistently reported that Arduino was manageable and approachable. This outcome was largely attributed to the use of block-coding interface programming tools such as mBlock with Gaduino. These interfaces replaced traditional coding syntax with intuitive visual elements, reducing the cognitive load typically associated with physical computing. Students' positive experiences underscore the role of thoughtfully designed tools and structured instruction in enhancing technological engagement in science classrooms.

Findings related to the third research question revealed several practical implications for teaching practice. First, students' decisions to adopt Arduino were sometimes driven by external motivations, such as preparing for university or aligning with perceived educational trends. Teachers should therefore help students understand the actual capabilities and constraints of Arduino, enabling them to select tools that best meet the needs of their research questions. In some cases, conventional equipment may be more suitable, and students should be prepared to evaluate all available options critically. Second, instruction should include guidance on how to assess sensor quality. Introducing students to sensor datasheets and helping them interpret specifications such as range, accuracy, and resolution would support more scientifically grounded decision-making. Third, educators should address students' questions about data reliability, particularly when using cloud-based platforms. Several students expressed concerns about not being able to view data directly. Teachers can respond by incorporating real-time display components or providing clear explanations of how data collection and transmission work. Improving students' understanding of these processes may increase their confidence in the technology and strengthen perceived usefulness.

Beyond these findings, this study makes several key contributions to the literature on science education and technology acceptance. First, by employing a qualitative methodology, it provides a deeply contextualized account that moves beyond the correlational findings of typical quantitative TAM studies, offering rich insights into the process of technology acceptance. Second, it challenges the prevailing narrative of Arduino as technically demanding for novices by demonstrating the transformative impact of visual block-coding tools like mBlock and Gaduino on Perceived Ease of Use. This finding has significant practical implications for curriculum design, suggesting a clear pathway to broader implementation. Third, it extends the application of TAM2 by illustrating how sociocultural factors, specifically the intense focus on university admissions in South Korea, manifest as powerful drivers of Subjective Norm and Image, influencing students' technology choices in ways not captured in other contexts.

While block-coding interfaces lowered barriers to engagement and improved ease of use, this raised broader pedagogical considerations. If the instructional goal is to emphasize scientific inquiry, visual programming tools may enable students to focus more effectively on experimental design and data interpretation. However, if curriculum goals also include computational thinking or coding literacy, there may be value in introducing students to text-based programming as well. A balanced approach that integrates both scientific and computational learning objectives could offer students a more comprehensive and meaningful STEM experience.

Despite its contributions, this study has limitations. the findings are not intended to be generalized across all settings. The study focused on a specific configuration involving Arduino UNO, mBlock, and Gaduino, without the inclusion of text-based programming. Variations in hardware, coding environments, or instructional design may result in different student experiences and outcomes. These limitations point to opportunities for further research. Future studies might explore whether students' perceptions of ease and usefulness are sustained when using text-based coding or alternative computing platforms. Comparative research examining different instructional approaches or school contexts would also provide valuable insights. Additionally, longitudinal studies could help determine whether positive early experiences with Arduino lead to long-term engagement with scientific inquiry and continued interest in STEM fields.

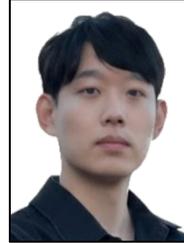

**Seok-Hyun Ga** was born in the Republic of Korea. He received Bachelor of Science degrees in Earth Science Education and Physics Education from Seoul National University, Seoul, Korea, in 2015, where he also earned the Master of Science in Education in 2017 and the Doctor of Philosophy in Science Education in 2021. He additionally completed a Bachelor of Science in Computer Science at Korea National Open University in 2021. He is currently a Visiting Assistant Professor at the National Taiwan Normal University, Taipei, Taiwan, and a Senior Lecturer at the Center for Educational Research at Seoul National University, Seoul, Korea. He has published multiple journal articles in IEEE Transactions on Learning Technologies, the Journal of Science Education and Technology, and the Journal of Chemical Education. His research focuses on action-oriented science education, Internet of Things applications in school science, science learning through virtual and augmented reality, and data analytics for science education.

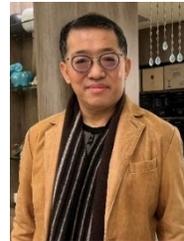

**Chun-Yen Chang (Fellow, IEEE)** was born in Taiwan. He received the B.S. degree in earth science from National Taiwan Normal University (NTNU), Taipei, Taiwan, and the M.Ed. and Ph.D. degrees in science education from The University of Texas at Austin, TX, USA.

He is currently a Chair Professor at the National Taiwan Normal University, where he also serves as Principal Investigator of the Institute for Research Excellence in Learning Sciences. He is also an Adjunct Professor at the State University of Malang, Indonesia, and a Concurrent Chair Professor at Chung Yuan Christian University, Taiwan. His research interests include science education, e-learning, interdisciplinary science learning, and science communication.

Prof. Chang is the Editor-in-Chief of *Eurasia Journal of Mathematics, Science and Technology Education*, *European Journal of Mathematics and Science Education*, *Educational Innovations and Emerging Technologies*, and *Applied System Innovation*. He also serves on the editorial boards of *Studies in Science Education*, *Learning, Media & Technology*, and *Journal of Science Education and Technology*. He has authored over 200 journal papers and received major national honors, including the Outstanding Research Award from Taiwan's National Science and Technology Council, the Academic Award for Science Education and the National Chair Professorship Award from the Ministry of Education, Taiwan.

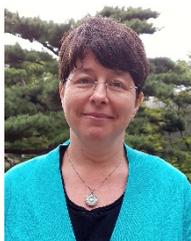 **Sonya N. Martin** is a Professor in the Departments of Science Education and Biology Education at Seoul National University, Republic of Korea. She received her B.A. in biology from Bryn Mawr College (USA), M.S.Ed. degrees in elementary and chemistry education from the University of Pennsylvania (USA), and a Ph.D. in science education from Curtin University (Australia). Her research focuses on equitable science teaching practices and teacher professionalization through practitioner-led research. Her recent work explores the integration of AI and chatbots in science classrooms. She has published widely and serves as Editor-in-Chief of *Asia-Pacific Science Education* (Brill).

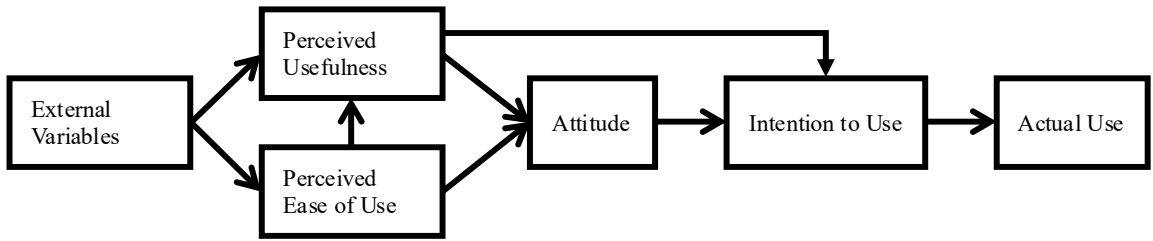

**Figure 1** *Original Technology Acceptance Model (TAM). [NO_PRINTED_FORM] [11]*

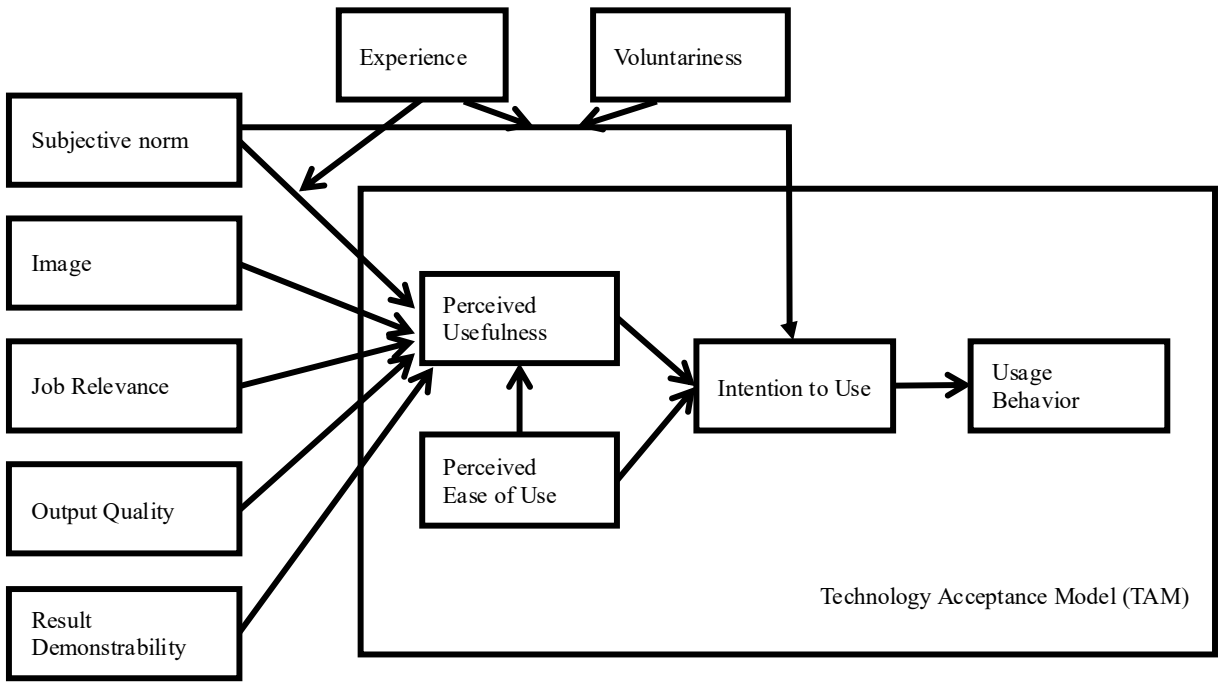

**Figure 2** *Technology Acceptance Model 2 (TAM). [NO_PRINTED_FORM] [22]*

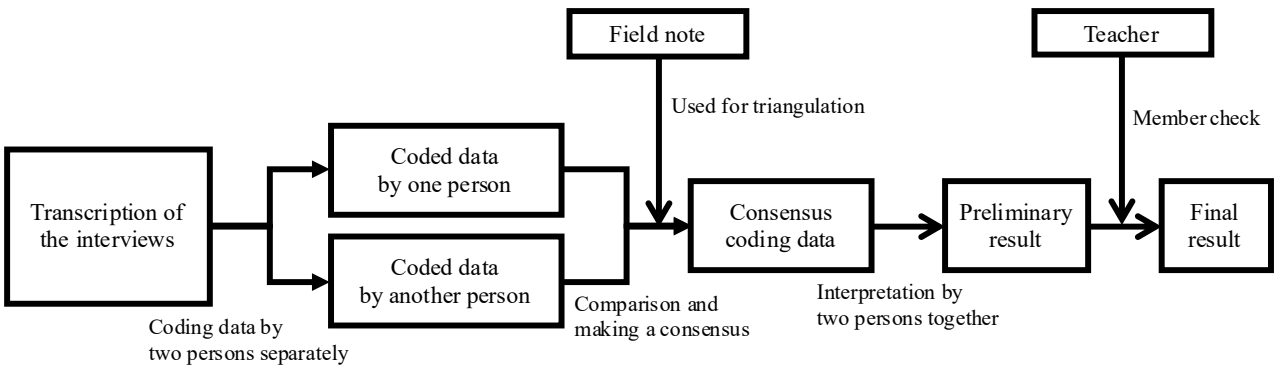

**Figure 3** Procedures of Data Analysis

**TABLE 1**

*After-School Research Class Schedule*

| Session | Main Activities and Topics |
|---|---|
| 1 | - Introduction to course objectives and schedule.<br>- Presentation of exemplary student-led projects to support topic development.<br>- Overview of the research study, including purpose, procedures, and voluntary participation. |
| 2 | - Introduction to Arduino using block-based coding to promote ease of use. |
| 3 | - Introduction to Internet of Things (IoT) concepts, including platform setup and Arduino data transmission.<br>- Demonstration of IoT-based data collection and visualization tools.<br>- Refinement of initial project ideas with teacher guidance; finalization of individual research topics. |
| 4 | - Identification and construction of custom measurement devices aligned with student projects. |
| 5 | - Finalization of detailed project plans.<br>- Outlining of specific steps for project implementation. |
| 6 | - Progress reporting and iterative feedback session.<br>- Continued documentation of project work.<br>- Instructional support provided by the teacher. |
| 7 | - Ongoing project adjustments and individual feedback (session continued from prior week). |
| 8 | - Formal presentation of student research outcomes. |
| 9 | - Structured reflection on overall project experience.<br>- Discussion of findings and challenges encountered. |

**TABLE 2**

*Components of the TAM2 model*

| Components | Description |
|---|---|
| PEU<br>Perceived Ease of Use | "The degree to which a person believes that using a particular system would be free of effort" [11, p. 320] |
| SN<br>Subjective Norm | "A person's perception that most people who are important to him think he should or should not perform the behavior in question" [28, p. 302]. |
| IMG<br>Image (and Social Influence) | "The degree to which use of an innovation is perceived to enhance one's … status in one's social system." [29, p. 195]. |
| JOB<br>Job Relevance | "An individual's perception regarding the degree to which the target system is applicable to his or her job" [22, p. 191]. |
| OQ<br>Output quality | "People will take into consideration how well the system performs those tasks" [22, p. 191]. |
| RD<br>Result demonstrability | "Tangibility of the results of using the innovation" [29, p. 203]. |
| Experience* | "The direct effect of subjective norm on intentions may subside over time with increased system experience" [22, p. 189]. |
| Voluntariness (and Compliance with Social Influence)* | "The extent to which potential adopters perceive the adoption decision to be non-mandatory" [30, p. 562]. |

**Note:** "Experience" and "Voluntariness" were not independently coded because they function as mediators affecting the influence of Subjective Norm on perceived usefulness or intention to use.